\documentstyle[prl,aps,epsf]{revtex}
\def\m#1{$#1$}

\def\tr{\;{\rm tr}\;}
\def\sgn{\;{\rm sgn}\;}
\newcommand{\beq}{\begin{equation}}
\newcommand{\eeq}{\end{equation}}

\newcommand{\beqs}{\begin{eqnarray}}
\newcommand{\eeqs}{\end{eqnarray}}

\newcommand{\half}{\frac{1}{2}}
\newcommand{\eps}{\epsilon}

\begin{document}
\title{\bf\large   Valence Parton Distribution Functions from Quantum ChromoDynamics  }

\author{
S. G. Rajeev\footnote{rajeev@pas.rochester.edu}\\
   {\it Department of Physics and Astronomy\\
 University of Rochester, Rochester, New York 14627} \\}

\maketitle

\abstract{ 

We show that two dimensional QCD can, to a good approximation,
describe the hadronic structure functions measured in Deep Inelastic
Scattering. We transform this theory  into a new form, Quantum
HadronDynamics (QHD) , whose semi-classical approximation is closer to
nature. The Baryon is then  a topological soliton, and its structure
function can be predicted  by a variational principle. This prediction
can be tested by comparison with measurements of neutrino scattering 
cross-sections.

}

%\vspace{3 cm}

{\it Keywords}: Structure Functions; Parton Model; Deep Inelastic
Scattering; Neutrino Scattering; QCD; Skyrme model; Quantum HadronDynamics.

{\it PACS }: 12.39Ki,13.60.-r, 12.39Dc,12.38Aw.

%\vspace{3 cm}

%\pagebreak

Perturbative QCD  allows us to determine the \m{Q^2}
dependence of the structure functions  of Deep 
Inelastic Scattering\cite{bjfeynman}. The initial condition for the
evolution of the DGLAP equations \cite{dglap}(i.e., the dependence on the Bjorken 
scaling variable \m{x_B} at an initial \m{Q^2}) cannot be determined
within the perturbative framework. As a result much effort has been
expended to extract these initial distributions by fits to data (See
e.g., \cite{mrst}.) We describe in this paper how to derive the
valence quark distributions of the proton ( in particular the
dependence on \m{x_B} at a low initial \m{Q^2}) from first
principles of QCD, by a succession of approximations.
It is possible to test the predictions 
against experimental data:  a comparison with the measurement of the
structure function \m{xF_3} in neutrino scattering will be done in a companion
 paper \cite{xf3}. The agreement is quite good, which confirms the theoretical 
framework  advocated in this paper. There are also some other approaches to 
this problem which are more numerical in character \cite{brodskyetal}

As we will explain below, Deep Inelastic Scattering can be explained
by the dimensional reduction of QCD to two spacetime dimensions. 
 The main idea is now to rewrite two
dimensional QCD in terms of operators that describe mesons rather than
quarks and gluons \cite{2dqhd}. This new formulation which I called
Quantum HadronDynamics (QHD) has the advantage that its semi-classical
approximation is quite close to nature: it corresponds to the large
\m{N_c} limit of QCD. Recall that the semi-classical approximation to
QCD itself is invalid except at short distances. In particular it
fails to explain the formation of hadrons. 

In QHD, the baryon appears as a toplogical soliton. Its structure
functions are determined by a variational principle, within the large
\m{N_c} limit. 
There is a natural variational ansatz 
which corresponds to the valence quark approximation. Within this
ansatz  we can even take care of the leading effect of \m{N_c} being
finite: it just amounts to restricting the range of momenta allowed for
partons. Thus we will be able to obtain a variational principle for
the valence quark distribution functions. A more detailed version of this 
argument can be found in \cite{istlect}.

Let us begin by recalling why Deep Inelastic Scattering can be
understood within a two dimensional framework \cite{bjfeynman}. An electron (or neutrino) scatters against a
proton (or other nucleus) with a space like momentum transfer of 
magnitude \m{q}, causing the target to disintegrate into many hadrons.
 The two Lorentz invariant variables that describe
this process can be chosen to be \m{Q^2=-q^2} and the Bjorken variable
 \m{x_B=-{q^2\over 2 p\cdot q}}. It is easy to check that \m{0<x_B<1}.
Deep inelastic scattering  is the
limiting case \m{Q^2\to \infty} keeping \m{x_B} fixed.  More precisely
\m{Q>>a^{-1}} where \m{a} is the characteristic size of the target. In
this limit, in the center of mass frame the target will look
`flattened out' to a pancake shape due to Lorentz contraction. To
first approximation, it can be thought of as having infinite extent in
the two tranverse directions while of finite extent \m{~a} in the
longitudinal direction. In other words the momenta of the constituents
of the hadron can be taken  to be zero in the transverse direction,
the corresponding fields are then  independent of the transverse
spatial co-ordinates. Thus the theory of strong interactions (QCD) can
be dimensionally reduced to \m{1+1} space-time dimensions in
describing deep inelastic scattering. We will use the  methods
of Ref.  \cite{2dqhd} to study  this two dimensional field
theory. This will give us the structure function as a function of \m{x_B}.

The effects of the transverse
momenta can then  be included as a perturbative correction. As
 emphasized  by Altarelli and Parisi \cite{dglap} this is precisely 
the meaning of the DGLAP evolution equations, which 
give the dependence of the structure functions on \m{Q^2}. (The
upper cut-off on the allowed transverse momentum is \m{Q}, which is how it 
is related to transverse momentum effects.) Since this part of the story 
is standard \cite{burascteq} we will not  discuss it here.

  In the above two dimensional approximation, the transversely
polarized gluons will appear as a pair of scalar fields.  Since we
don't  need the gluon structure functions for now, we can ignore these
transverse gluons. To leading order, the evolution of the 
 valence parton distributions decouple from the gluon distribution functions.
The longitudinal component of the gluon cannot be
ignored as it is responsible for the binding of the quarks into the
hadron. However, they don't have dynamical degrees of freedom and can
be eliminated using their equations of motion. 

Thus the action  of the two dimensional field theory describing the
Deep Inelastic Structure function is,
\beq
	S= {N_c\over {4g^2}}\int \tr F_{\mu\nu}F^{\mu\nu}d^2x+
\sum_{a=1}^{2N_f}\int \bar
q^{a\alpha}[-i\gamma\cdot\nabla+m_a]q_{a\alpha} d^2x.
\eeq  
Here, \m{\alpha=1,\cdots, N_c} is the color index and \m{a=1,\cdots, 2N_f} a
flavor index. Also, \m{q} is a two dimensional Dirac spinor ;
each four dimensional spinor splits into  a pair of these 
upon  dimensional reduction and hence the  two dimensional
theory has twice the number of flavors as  the four dimensional theory.
 In null 
co-ordinates (for more about the kinematics in the null co-ordinates, see 
\cite{istlect}) 
,and in the light--cone gauge \m{A_-=0}, we can eliminate
the remaining gluon degrees of freedom to get the hamiltonian
\beq
	H=\int dx\chi^{\dag a  i}  \half[\hat p+{m^2\over \hat p}]   \chi_{a i } -{1\over 2N}\alpha_1\int \half |x-y|
:\chi^{\dag a i}(x)\chi_{ a j}(x):
:\chi^{\dag b j}(y)\chi_{ b i}(y):dxdy
\eeq
where \m{q=\pmatrix{q_1\cr{1\over \surd 2}\chi}}. The upper component
\m{q_1} is not a propagating degree of freedom and has been eliminated
in terms of \m{\chi}. The quark field \m{\chi} satisfies
the canonical anti-commutation relations
\m{
	[\chi(x),\chi(y)]_+=\delta(x-y),\quad 
[\chi(x),\chi^\dag (y)]_+=0.
}
Also, the normal ordered product \m{:AB :} is defined with respect to
the vacuum 
\m{
\tilde\chi^{\dag}(p)|0>=0\;{\rm for}\; p<0,\quad 
\tilde\chi(p)|0>=0\;{\rm for}\; p>0.
}    

Now define the  color invariant variable
 \m{\hat M^a_b(x,y)={1\over N_c}:[\chi_{b\alpha}(x),\chi^{\dag a\alpha}(y)]:}
 which can be
 thought of as the field operator for a meson field. The space-time
 points \m{x,y} lie along a null line which is thought of as the
 initial value surface.

Now the entire theory can be described in terms of this color invariant
 variable. Within the subspace of color invariant states, \m{\hat M^a_b(x,y)}
 is a complete set of observables: the only operators that commute
 with them are multiples of the identity. This follows from the fact
 that \m{\hat M(x,y)} provide an irreducible (projective) unitary 
representation of
 the 
infinite dimensional unitary Lie algebra:
\beq
	\{\tilde {\hat M}^a_b(p,q),\tilde {\hat M}^c_d(r,s)\}=
{1\over N_c}\bigg(\delta_b^c 2\pi \delta(q-r)[\delta^a_d\sgn(p-s)+\tilde {\hat M}^a_d(p,s)]-\delta_d^a 2\pi \delta(s-p)[\delta^c_b\sgn(r-q)+\tilde {\hat M}^c_b(r,q)]\bigg). 
\eeq	
Here 
\m{
	\tilde {\hat M}^a_b(p,q)=\int \hat M^a_b(x,y)e^{ipx-iqy}dxdy.
}
Note that the commutators are of order \m{1\over N_c} so that the large
 \m{N_c} limit is a sort of classical limit: \m{1\over N_c} plays the
 role of \m{\hbar} in an ordinary field theory. 

In this classical limit the above commutators tend to Poisson
brackets of a  set of classical dynamical variables \m{M^a_b(x,y)}. The phase 
space of this classical 
dynamical system  must
be a homogenous symplectic manifold, this being the analogue of an
irreducible unitary representation. From the standard Kirillov theory
(adapted to infinite dimensions by Segal, \cite{presseg}) this phase
space is a co-adjoint orbit of the unitary group, the Grassmannian. It
is the set of all inifinite dimensional operators \m{M} with integral
kernel \m{M^a_b(x,y)} satisfying the nonlinear constraint
\m{
	[\eps+M]^2=1.
}
Here,
\m{\eps} is the Hilbert transform operator,
\m{
	\eps(x,y)=\int e^{ip(x-y)}\sgn(p){dp\over 2\pi}.
}
It is also possible to verify the identity above directly on color
singlet states as is shown in the appendix to Ref. \cite{sphqcd}.
Thus in the large \m{N_c} limit, our problem reduces to solving the equations of motion obtained from the hamiltonian 
\beq
	E[M]=-{1\over 4}\int [p+{\mu^2\over p}]\tilde M(p,p){dp\over
2\pi} +{\tilde g^2\over 8}\int M^a_b(x,y)M^b_a(y,x)|x-y|dxdy
\eeq
with the Poisson brackets
\beq
	{1\over 2i}\{M^a_b(x,y),M^c_d(z,u)\}=
\delta_b^c\delta(y-z)[\eps^a_d(x,u)+M^a_d(x,u)] -
\delta_d^a\delta(x-u)[\eps^c_b(z,y)+M^c_b(z,y)].
\eeq
The parameter \m{\mu_a^2} is related to the quark masses by a finite
renormalization: \m{\mu_a^2=m_a^2-{\tilde g^2\over \pi}}, where \m{m_a}
is the {\em current} quark mass. (Also, \m{\tilde M^a_b(p,q)=\int M^a_b(x,y)e^{-ipx+iqy}dxdy} is the Fourier transform.)

What kind of solution to this theory  represents the baryon? The quantity 
\m{
	B=-\half\int M^a_a(x,x)dx
}
can be shown to be an integer, a topological invariant \cite{2dqhd}. From the definition of
\m{M} in terms of \m{\chi,\chi^{\dag}} we can see that this is in fact
the baryon number. Thus the baryon is a topological soliton in this
picture: an idea originally proposed by Skyrme in quite a different
context, and revived by Balachandran et. al.  and by Witten et. al.
\cite{skyrme}. 
We seek a static solution (minimum of the  energy subject to
constraints) that has baryon number one. Again
from the definition in terms of the quark fields, we can see that
\m{-\half \tilde M^a_a(p,p)} represents the quark number density in momentum
space,when \m{p>0}. Similarly,  \m{\half \tilde M^a_a(-p,-p)} represents the 
anti-quark number density. It is convenient to assume a variational
ansatz of the separable (rank one) form 
\m{	\tilde  M^a_b(p,q)=-2\tilde \psi^{a}(p)\tilde \psi_b^*(q).
}
This satisfies the constraint if \m{\tilde \psi} is of norm one and of
positive momentum:
\m{
	\sum_a\int_0^\infty \tilde \psi_a(p)|^2{dp\over 2\pi}=1,\quad \tilde\psi_a(p)=0,\;{\rm for}\; p<0.
}
This variable satisfies the Poisson bracket relations
\m{
	\{\tilde\psi_a(p),\tilde\psi_b(q)\}=0,\quad
\{\tilde \psi_a(p),\tilde\psi^{*b}(q)\}=-i2\pi\delta_a^b\delta(p-q).
}
The energy becomes then,
\beq
	E_1(\psi)=\sum_a\int_0^\infty\half[p+{\mu^2\over
p}]|\tilde\psi_a(p)|^2{dp\over 2\pi} +
{\tilde g^2\over 2}\sum_{ab}\int |\psi_a(x)|^2|\psi_b(y)|^2 \half |x-y|dxdy.
\eeq
This variational ansatz corresponds to the valence quark
approximation: the anti-quark distributions are identicaly zero.
 In  forthcoming papers with V. John and G. S. Krishnaswami,
 we will  show that in the limit of zero current quark
mass the {\it exact } minimum of the energy functional is of this
separable form \cite{xf3,antiquark}. In fact 
deviations are small even for finite \m{m}; in the language of the
parton model, there is 
a less than one percent probability of
finding an anti-quark in the proton at low \m{Q^2}. The factorized
ansatz amounts to ignoring the anti-quarks.

So far we have worked  in the large \m{N_c} limit. What is the first order effect
of \m{N_c} being finite? If we stay within the valence parton  approximation as
we make \m{N_c} finite this can be studied by replacing the  Poisson
brackets of \m{\psi}  by canonical commutation relations: 
\beq
	[\hat{\tilde\psi}_a(p),\hat{\tilde\psi}_b(p')]=0=
[\hat{\tilde{\psi}}^{\dag a}(p),\hat{\tilde\psi}^{\dag b}(p')],
\quad
[\hat{\tilde\psi}_a(p),\hat{\tilde\psi}^{\dag b}(p')]=
{1\over N_c}2\pi\delta(p-p')\delta^b_a.
\eeq	
As usual, classical Poisson brackets go over to quantum commutation
relations, except that the role of \m{\hbar} is played by \m{1\over N_c}.
These commutation relations have a simple representation in terms of
bosonic creation-annihilation operators. The constraint on \m{\psi}
then becomes the condition that there be exactly \m{N_c} such bosons in
any allowed state. The large \m{N_c} limit is then like a thermodynamic
limit, in the canonical ensemble. 

What are these bosons? A moment's reflection will show that they are in
fact the valence partons: we have just given a derivation of the
valence parton model from a series of approximations on QCD. They
behave like bosons (rather than fermions) 
because we are not counting explicitly  the color quantum
number. The wavefunction of the system is completely anti-symmetric in
color (to make it color invariant) so that the Pauli principle
requires it to be symmetric in the remaining quantum
numbers. The null momentum of each parton is positive so each has to
be less than the total momentum \m{P}. The main effect of a finite
(but large) \m{N_c} is thus to require that \m{p<P}:
\m{
	\tilde\psi(p_1,\cdots,p_N)=0,\;\;{\rm if}\;\; p_i>P
}
which is in addition to the condition that \m{p_i>0}. This way we
derive exactly the model of interacting partons \cite{ipm}  as an approximation
to  two-dimensional QCD. As noted in that paper, we can make a mean
field approximation  keeping this condition in place to take into account of
the effect of finite \m{N_c}.

Thus we can determine the valence parton
distribution function if we can minimize the above energy functional \m{E_1(\psi)} 
subject to the conditions that 
\beq
	\tilde\psi(p)=0\;{\rm for }\; p<0\;{\rm and}\; p>P, \quad \int_0^P
|\tilde\psi(p)|^2{dp\over 2\pi}=1,\quad \int_0^P
p|\tilde\psi(p)|^2{dp\over 2\pi}=fP.
\eeq
Here \m{f} is the fraction of the momentum carried by the valence
partons.

This problem has been solved numerically \cite{ipm} as well as in a
variational approximation \cite{xf3}. The results can then be
compared to the experimental measurements of the \m{xF_3}
structure function. The agreement is  quite good, confirming our picture
of the structure of a hadron \cite{xf3}. In particular we have resolved the
apparent difference between the Skyrme  model  of the baryon
and the valence parton model: we have found a  topological  soliton 
 model that applies to high energy scattering from which an interacting
valence parton model can be derived. We can  derive more information 
 such as spin and flavor-dependent or
 anti-quark distributions functions \cite{antiquark}, and we hope to address these issues in
 later papers.

\noindent
Acknowledgement: I thank V. John and G. S. Krishnaswami for discussions. 
This research was supported in part by the US DOE grant DE-FG02-91ER40685.

\end{document}